 \documentclass[aps,prl,preprint,tightenlines,superscriptaddress,showpacs,byrevtex]{revtex4}
\usepackage{graphicx} 
\usepackage{xspace}
\usepackage{epsfig}
\usepackage{multirow}
\usepackage{dcolumn}  

\newcommand{\de}{\ensuremath{\Delta E}\xspace}

\newcommand{\acp}{\ensuremath{\mathcal{A}_{CP}}\xspace}

\newcommand{\bb}{\ensuremath{B \overline{B}}\xspace}

\def\myspecial#1{}                   
\def\calL{{\mathcal L}}

\def\Mbc{M_{\rm bc}}

\def\babar{{\sl BABAR}}

\graphicspath{{ps}}

\begin{document}


\preprint{\vbox{ \hbox{   }
                 \hbox{BELLE-CONF-0523}
                 \hbox{LP2005-158}
                 \hbox{EPS05-495} 
}}

\title{\quad\\[0.5cm] \hspace{-0.5cm}\Large
 Improved Measurements of Direct $CP$ Violation in \\
 $B \to K^+ \pi^{-}$,
 $K^+ \pi^{0}$
 and $\pi^+\pi^0$ Decays}

\affiliation{Aomori University, Aomori}
\affiliation{Budker Institute of Nuclear Physics, Novosibirsk}
\affiliation{Chiba University, Chiba}
\affiliation{Chonnam National University, Kwangju}
\affiliation{University of Cincinnati, Cincinnati, Ohio 45221}
\affiliation{University of Frankfurt, Frankfurt}
\affiliation{Gyeongsang National University, Chinju}
\affiliation{University of Hawaii, Honolulu, Hawaii 96822}
\affiliation{High Energy Accelerator Research Organization (KEK), Tsukuba}
\affiliation{Hiroshima Institute of Technology, Hiroshima}
\affiliation{Institute of High Energy Physics, Chinese Academy of Sciences, Beijing}
\affiliation{Institute of High Energy Physics, Vienna}
\affiliation{Institute for Theoretical and Experimental Physics, Moscow}
\affiliation{J. Stefan Institute, Ljubljana}
\affiliation{Kanagawa University, Yokohama}
\affiliation{Korea University, Seoul}
\affiliation{Kyoto University, Kyoto}
\affiliation{Kyungpook National University, Taegu}
\affiliation{Swiss Federal Institute of Technology of Lausanne, EPFL, Lausanne}
\affiliation{University of Ljubljana, Ljubljana}
\affiliation{University of Maribor, Maribor}
\affiliation{University of Melbourne, Victoria}
\affiliation{Nagoya University, Nagoya}
\affiliation{Nara Women's University, Nara}
\affiliation{National Central University, Chung-li}
\affiliation{National Kaohsiung Normal University, Kaohsiung}
\affiliation{National United University, Miao Li}
\affiliation{Department of Physics, National Taiwan University, Taipei}
\affiliation{H. Niewodniczanski Institute of Nuclear Physics, Krakow}
\affiliation{Nippon Dental University, Niigata}
\affiliation{Niigata University, Niigata}
\affiliation{Nova Gorica Polytechnic, Nova Gorica}
\affiliation{Osaka City University, Osaka}
\affiliation{Osaka University, Osaka}
\affiliation{Panjab University, Chandigarh}
\affiliation{Peking University, Beijing}
\affiliation{Princeton University, Princeton, New Jersey 08544}
\affiliation{RIKEN BNL Research Center, Upton, New York 11973}
\affiliation{Saga University, Saga}
\affiliation{University of Science and Technology of China, Hefei}
\affiliation{Seoul National University, Seoul}
\affiliation{Shinshu University, Nagano}
\affiliation{Sungkyunkwan University, Suwon}
\affiliation{University of Sydney, Sydney NSW}
\affiliation{Tata Institute of Fundamental Research, Bombay}
\affiliation{Toho University, Funabashi}
\affiliation{Tohoku Gakuin University, Tagajo}
\affiliation{Tohoku University, Sendai}
\affiliation{Department of Physics, University of Tokyo, Tokyo}
\affiliation{Tokyo Institute of Technology, Tokyo}
\affiliation{Tokyo Metropolitan University, Tokyo}
\affiliation{Tokyo University of Agriculture and Technology, Tokyo}
\affiliation{Toyama National College of Maritime Technology, Toyama}
\affiliation{University of Tsukuba, Tsukuba}
\affiliation{Utkal University, Bhubaneswer}
\affiliation{Virginia Polytechnic Institute and State University, Blacksburg, Virginia 24061}
\affiliation{Yonsei University, Seoul}
  \author{K.~Abe}\affiliation{High Energy Accelerator Research Organization (KEK), Tsukuba} 
  \author{K.~Abe}\affiliation{Tohoku Gakuin University, Tagajo} 
  \author{I.~Adachi}\affiliation{High Energy Accelerator Research Organization (KEK), Tsukuba} 
  \author{H.~Aihara}\affiliation{Department of Physics, University of Tokyo, Tokyo} 
  \author{K.~Aoki}\affiliation{Nagoya University, Nagoya} 
  \author{K.~Arinstein}\affiliation{Budker Institute of Nuclear Physics, Novosibirsk} 
  \author{Y.~Asano}\affiliation{University of Tsukuba, Tsukuba} 
  \author{T.~Aso}\affiliation{Toyama National College of Maritime Technology, Toyama} 
  \author{V.~Aulchenko}\affiliation{Budker Institute of Nuclear Physics, Novosibirsk} 
  \author{T.~Aushev}\affiliation{Institute for Theoretical and Experimental Physics, Moscow} 
% ...
\collaboration{The Belle Collaboration}
\noaffiliation

\begin{abstract}

We report an improved measurement of direct $CP$ violation
in the decay $B^0\to K^+\pi^-$,
and a search for $CP$ violation in the decays  
$B^+\to K^+\pi^0$ and $B^+\to \pi^+\pi^0$.
The measured $CP$ violating asymmetries are
${\cal A}_{CP}(K^+\pi^-) = -0.113\pm 0.022(\rm stat.)\pm 0.008(\rm syst.)$,
${\cal A}_{CP}(K^+\pi^0)   = 0.04\pm 0.04(\rm stat.)\pm 0.02(\rm syst.)$
and
${\cal A}_{CP}(\pi^+\pi^0) = 0.02\pm 0.08(\rm stat.)\pm 0.01(\rm syst.)$,
where the latter correspond to the intervals
$-0.03 < \acp(K^+ \pi^0) < 0.11$ 
and 
$-0.12 < \acp(\pi^+ \pi^0) < 0.15$
at 90\% confidence level.
These results are obtained from a data sample that contains 386
million $B\bar{B}$ pairs that was collected 
near the $\Upsilon(4S)$ resonance,
with the Belle detector at the KEKB asymmetric energy $e^+ e^-$
collider. All of the results are preliminary.

\end{abstract}


\maketitle
\tighten

{\renewcommand{\thefootnote}{\fnsymbol{footnote}}}
\setcounter{footnote}{0}

\normalsize

\newpage
In the Standard Model (SM) $CP$ violation arises via the interference of at
least two diagrams with comparable amplitudes but different $CP$ conserving
and violating phases. Mixing induced $CP$ violation in the $B$ sector has been
established in $b\to c\bar{c} s$ transitions \cite{2phi1,2beta}.
Direct $CP$ violation is expected to be sizeable in the $B$ meson system 
\cite{BSS}. The first experimental evidence for direct $CP$ 
violation was shown by Belle  for the decay mode $B^0\to \pi^+\pi^-$ 
\cite{PIPI}.
This result suggests large interference between tree and penguin 
diagrams and the existence of final state interactions \cite{FSI}. 
In addition,
both Belle \cite{belle_acp_250} and
{\babar} \cite{babar_acp_kpi_230}
have recently reported
evidence for direct $CP$ violation in the decay $B^0\to K^+\pi^-$.

 The partial rate $CP$ violating asymmetry is defined as: 
 \begin{eqnarray}
 \acp=\frac{N(\overline B \to \overline f)-N(B \to f)}
 {N(\overline B \to \overline f)+N(B \to f)},
 \end{eqnarray} 
 where $N(\overline B \to \overline f)$ is the yield for the
$\overline{B} \to K\pi/\pi\pi$ decay and $N(B \to f)$ denotes
that of the charge-conjugate mode. 
Theoretical predictions with different
approaches suggest that $\acp(K^+\pi^-)$ could be either positive  
or negative \cite{acpth}.  
Although there are large uncertainties related to  
hadronic effects in the theoretical predictions,  results for 
$\acp(K^+\pi^-)$ and $\acp(K^+\pi^0)$
are expected to have the same sign
and comparable magnitudes \cite{acpth}.
However, our previous measurements show that
${\acp}(K^+\pi^-)$ and ${\acp}(K^+\pi^0)$ are opposite in sign
(although $\acp(K^+\pi^0)$ is consistent with no asymmetry),
and their central values are found to deviate from each other
by $2.4\sigma$.
These findings are consistent with those reported
by {\babar} \cite{babar_acp_kpi_230,babar_acp_hpi0_230}.
It is suggested that the disagreement may be due to the contribution
of the electroweak penguin process or other mechanisms \cite{anom}.
Therefore, it is important to verify whether the discrepancy persists
with improved precision.
In this Letter, we report  $\acp$
measurements using 386 million $\bb$ pairs  collected
with the Belle detector at the KEKB $e^+e^-$ asymmetric-energy
(3.5 on 8~GeV) collider~\cite{KEKB} operating at the $\Upsilon(4S)$ resonance.

The Belle detector is a large-solid-angle magnetic
spectrometer that consists of a silicon vertex detector (SVD),
a 50-layer central drift chamber (CDC), an array of
aerogel threshold \v{C}erenkov counters (ACC),
a barrel-like arrangement of time-of-flight
scintillation counters (TOF),
and an electro-magnetic calorimeter
comprised of CsI(Tl) crystals (ECL) located inside 
a super-conducting solenoid coil that provides a 1.5~T
magnetic field.  An iron flux-return located outside of
the coil is instrumented to detect $K_L^0$ mesons and to identify
muons (KLM).
The detector is described in detail elsewhere~\cite{Belle}.
Two inner detector configurations were used. For the first sample 
of 152 million $\bb$ pairs (Set I), a 2.0 cm radius beampipe
and a 3-layer silicon vertex detector were used;
for the latter 234 million $\bb$ pairs (Set II),
a 1.5 cm radius beampipe, a 4-layer silicon detector
and a small-cell inner drift chamber were used\cite{Ushiroda}.
   
The $B$ candidate selection is the same as
that described in Ref.~\cite{btohh}.
Charged tracks are required to originate from the interaction point (IP).
Charged kaons and pions are identified using $dE/dx$
information and Cherenkov light yields in the ACC.
The $dE/dx$ and ACC information  are combined to form
a $K$-$\pi$ likelihood ratio, 
$\mathcal{R}(K\pi) = \mathcal{L}_K/(\mathcal{L}_K+\mathcal{L}_\pi)$,
where $\mathcal{L}_{K/\pi}$
is the likelihood of kaon/pion. Charged tracks with $\mathcal{R}(K\pi)>0.6$ are
regarded as kaons and $\mathcal{R}(K\pi)<0.4$ as pions. Furthermore,
charged tracks that are positively identified as electrons are rejected.
The $K/\pi$ identification efficiencies and misidentification
rates are determined from a sample of kinematically identified
$D^{*+}\to D^0\pi^+, D^0\to K^-\pi^+$ decays. Table \ref{tab:kid} shows the 
results. 
It is clear that the detection efficiency for
$K^-/\pi^+$ is greater  than for
$K^+/\pi^-$; this small efficiency bias will be corrected in the ${\cal A}_{CP}$ 
measurement.    

\begin{table}
\begin{center}
\caption{Performance of $K-\pi$ identification measured using 
$D^{*+}\to D^0\pi^+, D^0\to K^-\pi^+$ decays.}
\begin{tabular}{lcccc}
\hline\hline
   &  \multicolumn{2}{c}{Set I} & \multicolumn{2}{c}{Set II} \\ 
   & Eff. (\%) & Fake rate (\%) &  Eff. (\%) & Fake rate (\%) \\ \hline 
$K^+$   &$83.76\pm 0.18$ &$ 5.10\pm 0.12$ &$81.92\pm 0.15$ &$ 6.29\pm 0.10$ \\ 
$K^-$   &$84.76\pm 0.18$ &$ 5.69\pm 0.12$ &$82.79\pm 0.14$ &$ 6.71\pm 0.10$ \\
$\pi^+$ &$91.24\pm 0.15$ &$10.72\pm 0.15$ &$89.88\pm 0.12$ &$12.28\pm 0.12$ \\
$\pi^-$ &$90.53\pm 0.15$ &$10.08\pm 0.15$ &$89.08\pm 0.13$ &$11.83\pm 0.12$ \\ \hline
\end{tabular}
\label{tab:kid}
\end{center}
\end{table}

Candidate $\pi^0$ mesons are reconstructed by combining two photons with
invariant mass between 115 MeV/$c^2$ and 152 MeV/$c^2$, which corresponds to
$\pm2.5$ standard deviations. Each photon is required to have a minimum
energy of 50 MeV in the barrel region ($32^\circ < \theta_\gamma < 129^\circ$)
or 100 MeV in the end-cap region ($17^\circ < \theta_\gamma < 32^\circ$ or
$129^\circ < \theta_\gamma < 150^\circ$), where $\theta_\gamma$ denotes the
polar angle of the photon with respect to the beam line.
To further reduce the combinatorial background, $\pi^0$ candidates
with small decay angles ($\cos\theta^* >0.95$) are rejected, where
$\theta^*$ is the angle between a $\pi^0$ boost direction from the
laboratory frame and its $\gamma$ daughters in the $\pi^0$ rest frame.

Two variables are used to identify $B$ candidates: the beam-constrained mass,
$M_{\rm bc} =  
\sqrt{E^{*2}_{\mbox{\scriptsize beam}} - p_B^{*2}}$, and the energy difference,
$\Delta E = E_B^* - E^*_{\mbox{\scriptsize beam}}$, where 
$E^*_{\mbox{\scriptsize beam}}$ is the beam energy and $E^*_B$ and $p^*_B$ are
the reconstructed energy and momentum of the $B$ candidates in the
center-of-mass (CM) frame. Events with 
$M_{\rm bc} > 5.20$ GeV/$c^2$ and $-0.3~{\rm GeV} < \Delta E < 0.5~{\rm GeV}$
are selected for the final analysis. 

The dominant background is from $e^+e^- \to q\bar q ~( q=u,d,s,c )$ continuum
events. To distinguish the signal from the jet-like continuum background,
event topology variables and $B$ flavor tagging information are employed.
We combine a set of modified Fox-Wolfram moments \cite{pi0pi0} into a
Fisher discriminant. The probability density functions (PDF) for this
discriminant, and the cosine of the angle between the $B$ flight
direction and the $z$ axis, are obtained using signal and continuum
Monte Carlo (MC) events. These two variables are then combined to form
a likelihood ratio
$\mathcal{R} = {\calL}_s/({\calL}_s + {\calL}_{q \bar{q}})$,
where ${\calL}_{s (q \bar{q})}$ is
the product of signal ($q \bar{q}$) probability densities. Additional
background discrimination is provided by $B$ flavor tagging.
The  standard Belle flavor tagging algorithm \cite{tagging} gives two
outputs: a discrete variable indicating the flavor of the tagging $B$
and the MC-determined dilution factor $r$,
which ranges from zero for no flavor information to unity for unambiguous
flavor assignment. An event that contains
a lepton ($r$ close to unity) is more likely to be a $B \overline B$ event
so a looser $\mathcal{R}$ requirement can be applied. We divide the data 
into $r>0.5$ and $r<0.5$ regions.
The continuum background is reduced by applying a selection requirement on 
 $\mathcal{R}$ for events in each $r$ region of Set I and Set II 
according to the figure-of-merit defined as
$N_s^{exp}/\sqrt{N_s^{exp}+N_{q\bar{q}}^{exp}}$, where $N_s^{exp}$ denotes
the expected signal yields based on our previous branching fraction
measurements \cite{btohh} and $N_{q\bar{q}}^{exp}$ denotes the expected
$q\bar q$ yields from sideband data ($M_{\rm bc}<5.26$ GeV/$c^2$).
A typical requirement suppresses 92--99\% of the continuum background while
retaining 48--67\% of the signal.

Backgrounds from $\Upsilon(4S) \to B\overline B$ events are investigated using
a large MC sample. After the $\mathcal{R}$  requirement, 
we find a small charmless three-body background at low $\Delta E$, and 
reflections from other $B^0\to \pi^+\pi^-$ decays due to $K$-$\pi$ 
misidentification.
   
The signal yields are extracted by applying unbinned two dimensional
maximum likelihood (ML) fits to the ($M_{\rm bc}$, $\Delta E$)
distributions of the $B$ and $\overline B$ samples.
The likelihood for each mode is defined as
\begin{eqnarray}
\mathcal{L} & = & {\rm exp}\; (-\sum_{s,k,j} N_{s,k,j}) 
\times \prod_i (\sum_{s,k,j} N_{s,k,j} {\mathcal P}_{s,k,j,i}) \\ 
\mathcal{P}_{s,k,j,i} & = &\frac{1}{2}[1- q_i \acp{}_j ]
P_{s,k,j}(M_{{\rm bc}i}, \Delta E_i),  
\end{eqnarray}
where $s$ indicates Set I or Set II, $k$ distinguishes events in the $r<0.5$ 
or $r>0.5$ regions, $i$ is the identifier of
the $i$-th event, $P(M_{\rm bc}, \Delta E)$ is the two-dimensional PDF,
$q$ indicates the reconstructed $B$ meson flavor,
$B (q=+1)$ or $\overline{B} (q=-1)$,
$N_j$ is the number of events for the category $j$, 
which, in turn, corresponds to either signal, $q\bar{q}$ continuum,
a reflection due to $K$-$\pi$ misidentification, or
background from other charmless three-body $B$ decays.  

The yields and asymmetries for the signal and backgrounds
are allowed to float in all modes. 
Since the $K^+\pi^0$ and
$\pi^+\pi^0$ reflections are difficult to distinguish with $\Delta E$
and $M_{\rm bc}$, we fit these two modes simultaneously with a fixed
reflection-to-signal ratio based on the measured $K$-$\pi$ identification 
efficiencies
and fake rates. All the signal PDFs ($P(M_{\rm bc},\Delta E)$) are obtained
using MC simulations based on the Set I and Set II detector configurations. 
The same signal PDFs are used for events in two different $r$ 
regions. No strong correlations between 
$M_{\rm bc}$ and $\Delta E$ are found for the  $ B\to K^+\pi^-$
signal. Therefore, its PDF is modeled by a product of a
single Gaussian for $M_{\rm bc}$ and a double Gaussians for $\Delta E$.
For the modes with a $\pi^0$ meson in the final state, there are correlations
between $M_{\rm bc}$ and $\Delta E$ in the tails of the signals; hence,
their PDFs are described by smoothed two-dimensional histograms.
Discrepancies between the signal peak positions in data and MC are calibrated
using $B^+ \to \overline{D}{}^0\pi^+$ decays, where the
$\overline{D}{}^0 \to K^+\pi^-\pi^0$ sub-decay is used for the modes with a
$\pi^0$ meson while $\overline{D}{}^0\to K^+\pi^-$ is used for the $K^+\pi^-$
mode.
The MC-predicted $\Delta E$ resolutions are verified using the
invariant mass distributions of high momentum $D$ mesons. The decay mode 
$\overline{D}{}^0\to K^+\pi^-$ is used for $B^0\to K^+ \pi^-$, 
and
$\overline{D}{}^0\to K^+\pi^-\pi^0$ for the modes with a $\pi^0$ in the final
state. 
The parameters that describe the shape of the signal PDFs are
fixed in all of the fits.

The continuum background in $\de$ is described by a first or second order
polynomial while the $\Mbc$ distribution is parameterized by an
Argus function $f(x) = x \sqrt{1-x^2}\;{\rm exp}\;[ -\xi (1-x^2)]$, where
$x$ is $\Mbc$ divided by half of the total center of mass energy.
The continuum PDF is the product of an Argus function and a polynomial, where parameters 
$\xi$ and the coefficients of the polynomial are free parameters.
These free parameters are $r$-dependent.
A large MC sample is used to investigate the background 
from charmless $B$ decays and a smoothed two-dimensional histogram
is taken as the PDF.
The functional forms of the PDFs are the same
for the $B$ and $\overline{B}$ samples.

The efficiency of particle identification is slightly
different for positively and negatively charged particles;
consequently,
the parameter ${\cal A}_{CP}$ in Eq.3 becomes
an effective partial rate asymmetry. 
For the $K^+\pi^0$ and $\pi^+\pi^0$ modes, this raw asymmetry
can be expressed as:
\begin{eqnarray}
 {\cal A}^{\rm raw}_{CP} =
  \frac{{\cal A}_\epsilon + {\cal A}_{CP}}{1 + {\cal A}_\epsilon {\cal A}_{CP}},
\end{eqnarray} 
where ${\cal A}_{CP}$ is the true partial rate asymmetry and the efficiency asymmetry 
${\cal A}_\epsilon$ is the efficiency difference
between $K^-(\pi^+$) and $K^+(\pi^-$)
divided by the sum of their efficiency. 
The situation is more complicated for 
the $K^+\pi^-$ mode because, in addition to the bias due to the efficiency 
difference between $K^-\pi^+$  and $K^+\pi^-$, a $K^-\pi^+$ signal event 
can be doubly misidentified as a $K^+\pi^-$ candidate and dilute ${\cal A}_{CP}$.
The efficiency asymmetry results in a ${\cal A}_{CP}$ bias of +0.01,
while the small dilution factor due to double misidentification reduces the 
$\acp$ by a factor of 0.98.
 
\begin{figure*}
\hspace{-1.0cm}
\epsfig{file=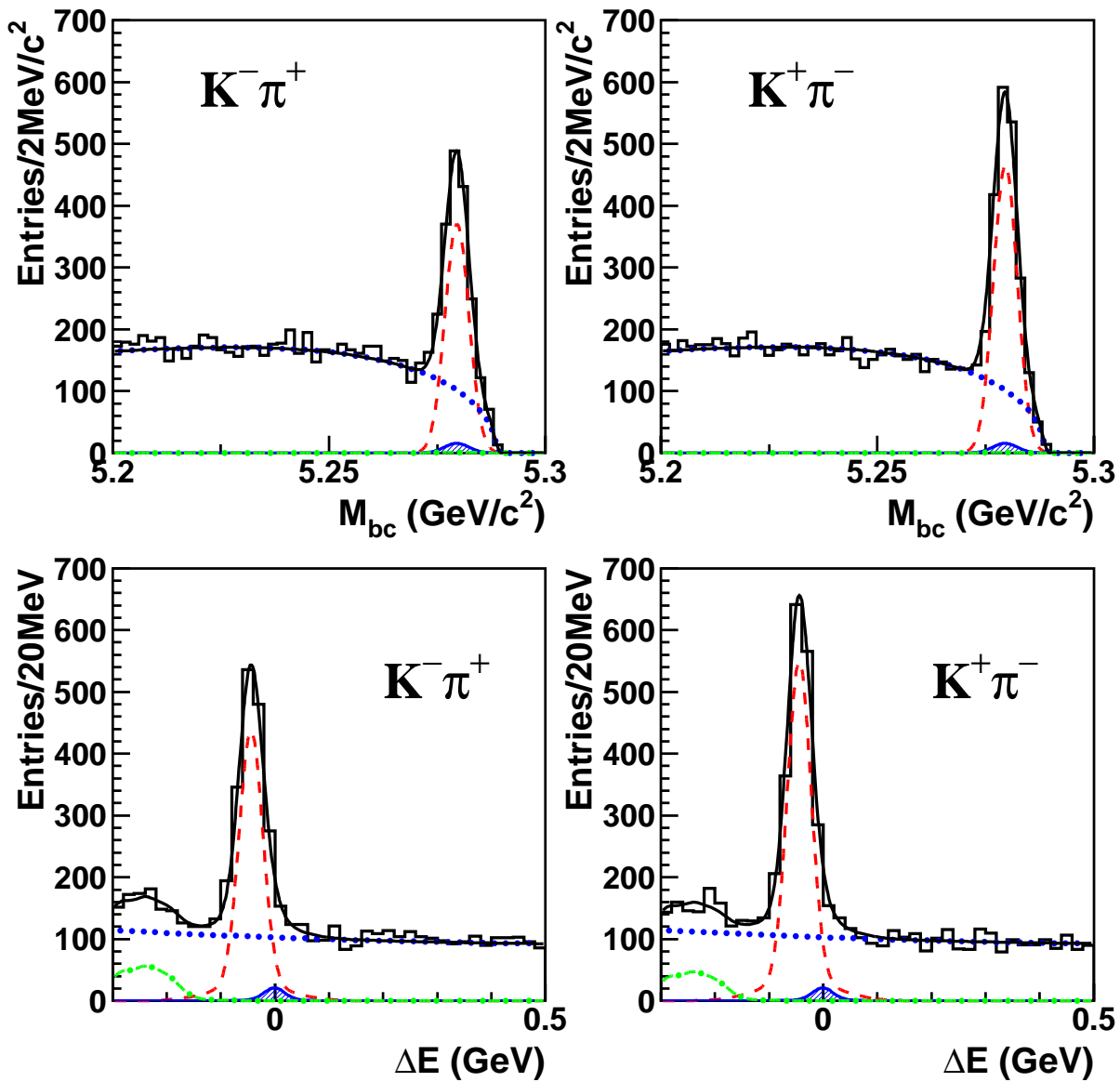,width=6.5in} 
\caption{$M_{\rm bc}$ (top) and $\Delta E$ (bottom) distributions for
${\overline B} ^0\to K^-\pi^+$ (left) and $B^0\to K^+\pi^-$ (right)
candidates. The histograms
represent the data, while the curves represent the various components from
the fit: signal (dashed), continuum (dotted), three-body B decays
(dash-dotted), background from  misidentification (hatched),
and sum of all components (solid).}
\label{fig:kpi}
\end{figure*}

\begin{figure*}
\hspace{-1.0cm}
\epsfig{file=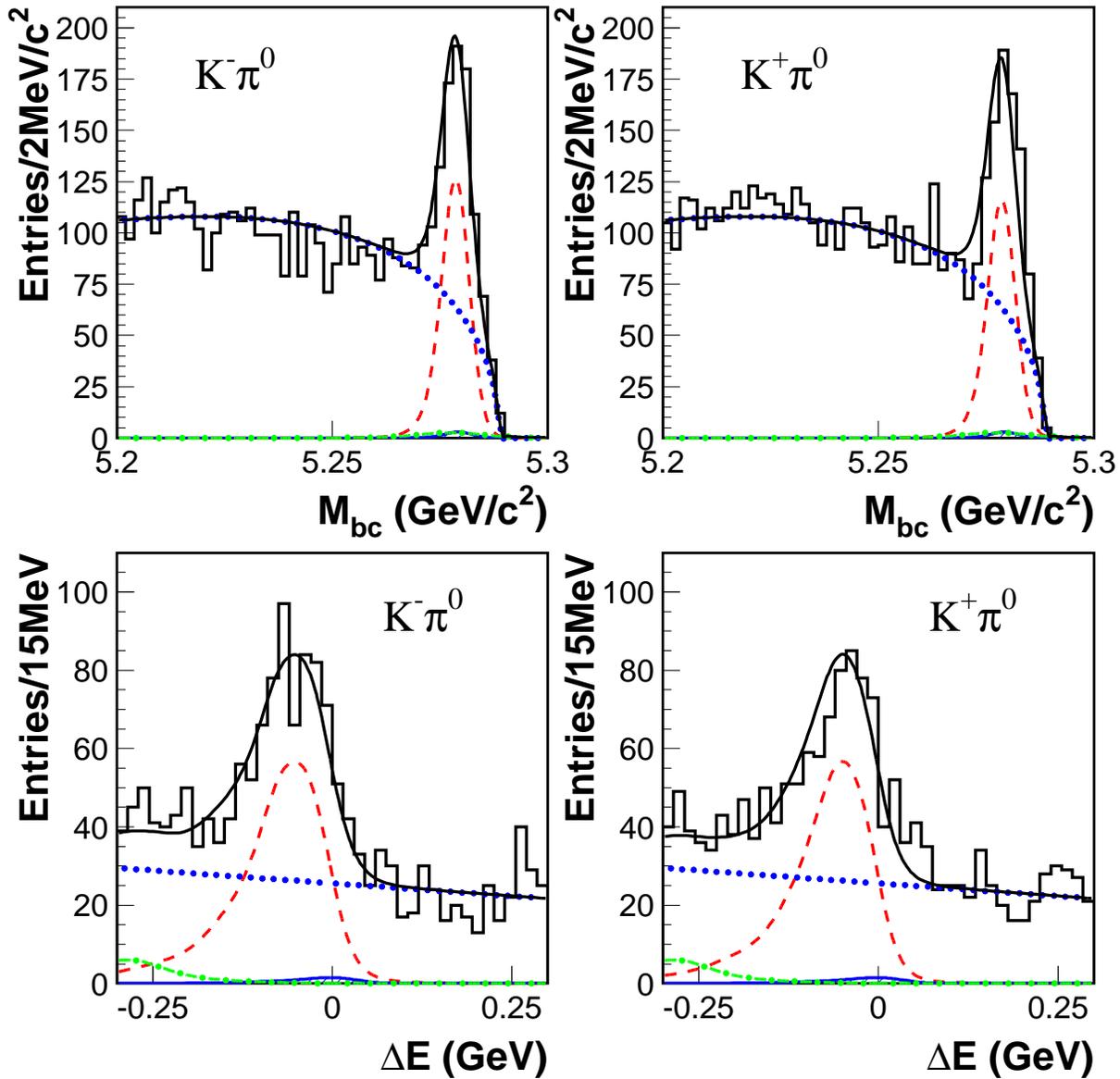,width=6.5in} 
\caption{$M_{\rm bc}$ (top) and $\Delta E$ (bottom) distributions for
$B^-\to K^-\pi^0$ (left) and $B^+\to K^+\pi^0$ (right) candidates.
The curves are described in the caption of Fig.\ref{fig:kpi}.}
\label{fig:kpi0}
\end{figure*}
   
Table \ref{tab:acp} shows the signal yields and $\acp$ values for each mode.
The asymmetries for the background components are consistent with zero within
errors.
Projections of the fits are shown in Figs.1-3.
The systematic errors from fitting are estimated  
from the deviations in $\acp$ after varying each parameter of the 
signal PDFs by 1 standard deviation. The uncertainty in modeling 
the three-body background is studied by excluding the low $\Delta E$ region
($<-0.12$ GeV) and repeating the fit. 
Systematic uncertainty due to particle identification is estimated
by repeating the fit after varying the $K/\pi$ efficiencies and fake rates
by 1 standard deviation. 
At each step, the deviation in $\acp$ is added in
quadrature to provide the systematic errors, which are   
less than 0.01 for all modes. A possible bias from the fitting procedure is  
checked in MC and a bias due to the $\mathcal{R}$ requirement is investigated 
using the $B^+\to \overline{D}{}^0\pi^+$ samples. No significant bias is 
observed.  
The systematic uncertainties due to the detector bias are tested
using the fit results for the continuum background listed in Table \ref{tab:acp}. 
We find a small background  asymmetry dependence on the ${\cal R}$ requirement
for the $K^+\pi^-$ mode,
and assign the uncertainty from the fit result of the 
$B^+\to \overline{D}{}^0\pi^+$($\overline{D}{}^0 \to K^+\pi^-$)
sample ($\pm0.007$)  as the systematic uncertainty due to detector bias.
The final systematic errors are then obtained
by quadratically summing the
errors due to the detector bias and the fitting systematics. 

\begin{table}
\begin{center}
\caption{Fitted signal yields, $\acp$ results, and background asymmetries for
 individual modes.}
\begin{tabular}{lrcc}
\hline\hline
~Mode~ & ~~~Signal Yield & $\acp$ & Bkg $\acp$\\
\hline
~$K^\mp\pi^\pm$ &$3026\pm 63$ &$-0.113\pm 0.022\pm 0.008$~ &$-0.001\pm 0.004$\\
~$K^\mp\pi^0$   &$1084\pm 45$ &$+0.04 \pm 0.04 \pm 0.02$   &$-0.02 \pm 0.01$ \\
~$\pi^\mp\pi^0$ &$ 454\pm 36$ &$+0.02 \pm 0.08 \pm 0.01$   &$-0.01 \pm 0.01$ \\
\hline\hline
\end{tabular}
\label{tab:acp}
\end{center}
\end{table}

The partial rate asymmetry $\acp(K^+\pi^-)$ is found to be
$-0.113\pm 0.022\pm 0.008$.
The significance including the effect of systematic uncertainty
is 4.97$\sigma$.
This result supersedes our previous measurement \cite{belle_acp_250} and
remains consistent with the value reported by {\babar},
$\acp(K^+\pi^-) = -0.133 \pm 0.030 \pm 0.009$ \cite{babar_acp_kpi_230}.
The observed $\acp(K^+\pi^0)$ value is consistent with zero
at the current level of statistical precision.
The difference between the results for $\acp(K^+\pi^-)$ and $\acp(K^+\pi^0)$
persists;
their central values differ by
$3.1 \sigma$.
This suggests a possible contribution from the electroweak penguin process
or other mechanisms \cite{anom}.
No evidence of
direct $CP$ violation is observed in the decay $B^+ \to \pi^+\pi^0$.
We set  90\% C.L. intervals: 
$ -0.03 < \acp(K^+ \pi^0) < 0.11$ 
and 
$ -0.12 < \acp(\pi^+ \pi^0) < 0.15$.
All of the above results are preliminary.

 %
 %

\begin{figure*}
\hspace{-1.0cm}
\epsfig{file=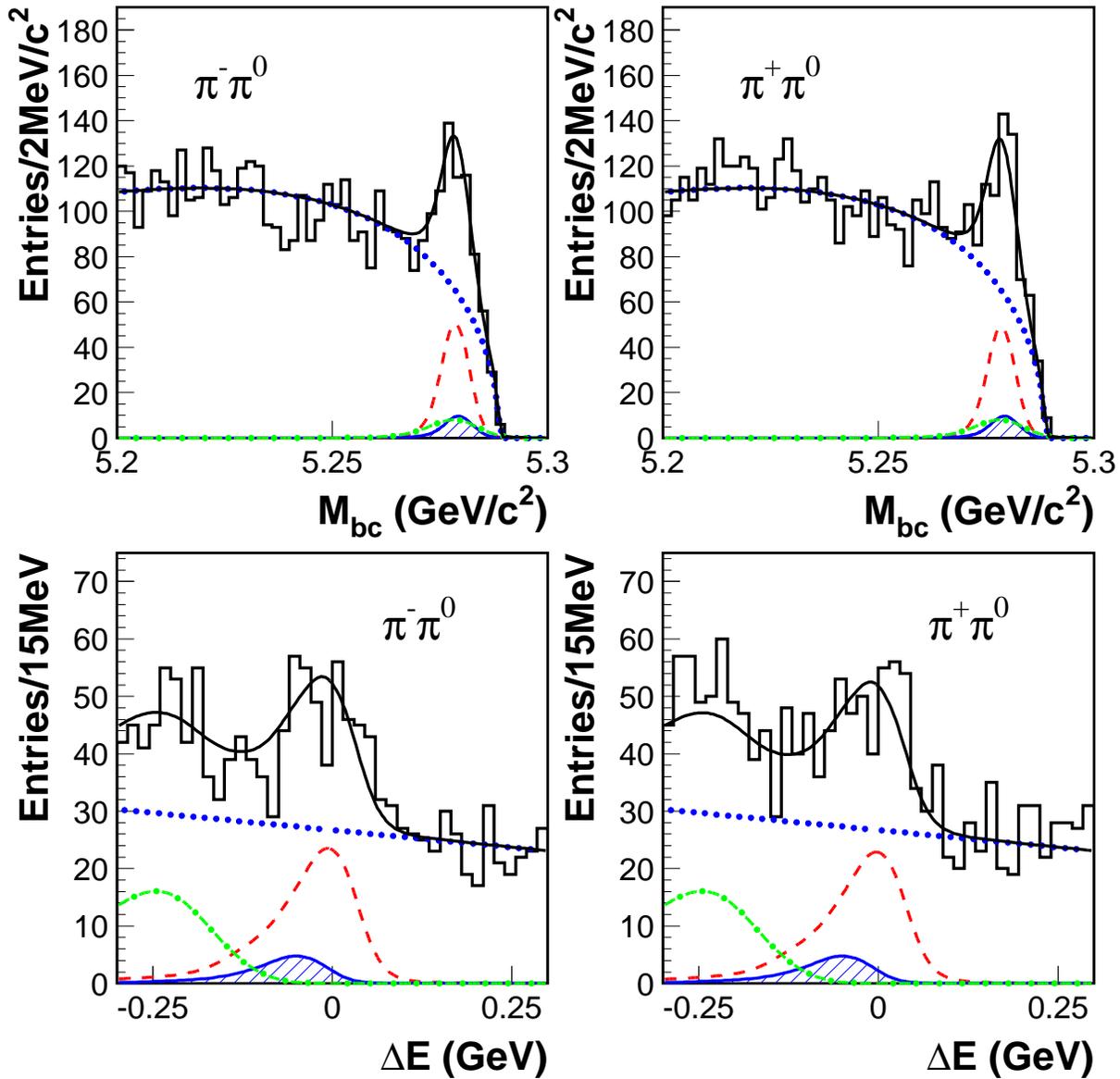,width=6.5in} 
\caption{$M_{\rm bc}$ (top) and $\Delta E$ (bottom) distributions for
$B^-\to \pi^-\pi^0$ (left) and $B^+\to \pi^+\pi^0$ (right)
candidates.
The curves are described in the caption of Fig.\ref{fig:kpi}.}
\label{fig:pipi0}
\end{figure*}

We thank the KEKB group for the excellent operation of the
accelerator, the KEK cryogenics group for the efficient
operation of the solenoid, and the KEK computer group and
the National Institute of Informatics for valuable computing
and Super-SINET network support. We acknowledge support from
the Ministry of Education, Culture, Sports, Science, and
Technology of Japan and the Japan Society for the Promotion
of Science; the Australian Research Council and the
Australian Department of Education, Science and Training;
the National Science Foundation of China under contract
No.~10175071; the Department of Science and Technology of
India; the BK21 program of the Ministry of Education of
Korea and the CHEP SRC program of the Korea Science and
Engineering Foundation; the Polish State Committee for
Scientific Research under contract No.~2P03B 01324; the
Ministry of Science and Technology of the Russian
Federation; the Ministry of Higher Education, 
Science and Technology of the Republic of Slovenia;  
the Swiss National Science Foundation; the National Science Council and
the Ministry of Education of Taiwan; and the U.S.\
Department of Energy.

\end{document}